\documentclass[twocolumn]{article}
\usepackage{authblk}
\usepackage{amsmath,amssymb,amsfonts,amsthm}
\usepackage{algorithmic}
\usepackage{color}
\usepackage{graphicx}
\usepackage{mathtools}
\usepackage{textcomp}
\usepackage{xcolor}
\usepackage{relsize}
\usepackage[colorlinks,allcolors=blue]{hyperref}
\usepackage{float}
\usepackage{dblfloatfix}
\usepackage{geometry}
\title{Automatic Query Optimization for Retrieving Traffic Tweets}
\author[1]{Emory Hufbauer}
\affil[1]{Researcher, Department of Computer Science, University of Kentucky}
\author[2]{Hana Khamfroush}
\affil[2]{Assistant professor, Department of Computer Science, University of Kentucky}

\begin{document}
\date{}
\vspace{-15mm}
\maketitle

\begin{abstract}
Twitter, like many social media and data brokering companies, makes their data available through a search API (application programming interface). In addition to filtering results by date and location, researchers can search for tweets with specific content with a boolean text query, using {\it AND}, {\it OR}, and {\it NOT} operators to select the combinations of phrases which must, or must not, appear in matching tweets. This boolean text search system is not at all unique to Twitter and is found in many different contexts, including academic, legal, and medical databases, however it is stretched to its limits in Twitter's use case because of the relative volume and brevity of tweets. In addition, the semi-automated use of such systems was well studied under the topic of Information Retrieval during the 1980s and 1990s, however the study of such systems has greatly declined since that time. As such, we propose updated methods for automatically selecting and refining complex boolean search queries that can isolate relevant results with greater specificity and completeness. Furthermore, we present preliminary results of using an optimized query to collect a sample of traffic-incident-related tweets, along with the results of manually classifying and analyzing them.
\end{abstract}

\section{Introduction}
Mining data from social media sites such as Twitter is a topic of interest in many research areas. However, efficiently finding relevant posts (e.g. tweets) among the billions that are produced each year remains a challenge. For our application, we are interested in identifying tweets which contain information relevant to traffic prediction in a certain geographic area. In addition to presenting a new genetic algorithm technique for query optimization, we also present some initial results in surveying the varieties, volumes, and sources of available traffic-related tweets. Genetic algorithms have been investigated for information retrieval applications before, for example by \cite{tamine2003multiple}, but not in a contemporary context.

\subsection{Twitter Queries}
The API accepts some number of request tokens (with a cost of approximately \$1 each) and a query, and returns up to 500 tweets matching the query per token submitted, or as many as exist in Twitter's database. The query consists of two parts: a metadata query and a content query. The metadata query filters tweets based on data such as the date they were posted, the account name or account location, or whether or not it is geotagged, or contains an image, or is a retweet. The content query selects tweets based on their text content. It is essentially a Boolean search query composed of words or phrases combined with the operators AND, OR, and NOT.

\subsection{Problem Definition}

The total length of the query is limited in practice to 1024 characters, but in general queries with this syntax have an arbitrarily deep tree structure. To reduce complexity, we are considering a smaller but equivalent subset of queries, called clause-structured queries: Those which have depth three, and strict operator order $\text{AND} < \text{OR} < \text{NOT}$. A clause-structured query is a set of clauses; each clause is a set of text phrases (each potentially subject to a NOT operator). To construct a Boolean search query from a clause-structured query, we use the OR operator to combine the phrases within each clause and the AND operator to combine the clauses together. In this way, a tweet will match a clause-structured content query iff it contains at least one positive (or does not contain one negative) phrase from each clause. It follows that a clause containing only a single negative phrase excludes all tweets containing that phrase.

\begin{gather*}
    \{ \{\text{phrase1}, \text{phrase2}\}, \{ -\text{phrase3} \} \} \\ \Downarrow \\
    \text{(phrase1 OR phrase2) AND (NOT phrase3)}
\end{gather*}

By applying De Morgan's laws, we can rearrange any query into this form, analogously to the conjunctive normal form to which all simple Boolean expressions can be reduced.\cite{losee1988integrating} This simplification allows our query optimization approaches to consider all possible queries within a smaller search space.

A query $Q$ matches a tweet $t$ ($Q(t)$ is true) iff each clause $q_i$ matches:
\begin{equation*}
    Q(t) = \bigwedge\limits_{q_i \in Q} q_i(t)
\end{equation*}
\begin{equation*}
    q_i(t) = \bigvee\limits_{(\phi, s_{\phi}) \in \Phi(q_i)} \begin{cases}
    \,\,\, s_{\phi},       & \phi \in \Phi(t) \\
    \neg s_{\phi},  & \phi \notin \Phi(t)
    \end{cases}
\end{equation*}
Here, $\Phi(q_i)$ is the set of phrases $\phi$ in $q_i$, and their signs, $s_{\phi}$. We can simplify the conditional logic using the {\it XNOR} or boolean equality operator, and combine with the definition of $Q(T)$:
\begin{equation*}
    Q(t) = \mathsmaller{\bigwedge\limits_{\mathsmaller{q_i \in Q}}} \,\, \mathlarger{\bigvee}\limits_{\mathclap{\mathsmaller{(\phi, s_{\phi}) \in \Phi(q_i)}}} \,\, s_{\phi} \odot (\phi \in \Phi(t))
\end{equation*}
\begin{center}
\begin{tabular}{ c|c c }
 $p \odot q$  & $\!\!\! \neg p$ & $p$ \\ \hline
 $\neg q$  & 1               & 0 \\
 $\,\,\,q$ & 0               & 1    
\end{tabular}
\end{center}

In general, finding the appropriate sets of phrases to construct a query which will match all relevant tweets and as few irrelevant ones as possible, is a subtle and difficult (NP-hard) task. This has been proven before in the literature, for example by Safari et al \cite{safari2019optimizing} in 2019, who demonstrated a reduction from the Set Max-coverage problem to a formulation of the query optimization problem described above. However, that formulation was limited in that it considered only simple queries, with a single clause and no negative phrases. By extending the mathematical descriptions above, we plan to prove demonstrate a new two-way equivalence between the complete query optimization problem and the Max-Satisifiability problem. If successful, this will both prove that Boolean search query optimization is actually NP-complete, and provide a new potential method for solving it leveraging existing SAT solvers.

\section{Algorithm}
Our genetic algorithm approach attempts to evolve a population of queries which minimize a loss function based on their false negative and false positive rates: $\textit{loss} = \frac{(f_p + \epsilon_{f_p}) (f_n + \epsilon_{f_n})}{(1+\delta_{f_p}-f_p)(1+\delta_{f_n}-f_n)}$, where $f_n$ and $f_p$ are the false negative and positive rates of the query, respectively; $\epsilon$ and $\delta$ parameters control the bias of the loss function towards selective vs complete queries. Because a thorough genetic search may require millions of queries to be evaluated, testing them directly against the Twitter API is impractical. Instead, we propose a dual-evaluation scheme: Every several generations, a single high-fitness (low loss) query is selected from the population, and sent to the Twitter API. The returned tweets are then analyzed, processed into sets of n-grams, and added to a local database. The other queries in the population are then evaluated against this local database in parallel. This allows many tweets to be approximately evaluated with a single API call.

The training database for our query optimizer consists of two components: a {\it vocabulary index}, and a set of {\it tweet vectors}. The vocabulary index is a list containing every $(n<4)$-gram which appears in more than one tweet from our dataset, ordered by frequency. In this way, each word or n-gram in the dataset is assigned an index, and more common words have lower indices. Let the size of the vocabulary index, which in practice is typically in the low millions, be denoted by the variable $n$. Meanwhile, each tweet vector corresponds to a single tweet; it has length $n$, and each element is a Boolean variable indicating whether the $(i < n)$th n-gram in the vocabulary index is present in that tweet. These boolean vectors are then bit-packed into vectors of 64-bit integers. In this way, each tweet may be represented in a compact, numerical way, which enables efficient evaluation.

Queries are then considered as sequences of integers. Positive integer elements correspond to members of the vocabulary index, while negative elements correspond to their negations. Zero is used a special separator character, indicating the end of a clause. We define five mutation and two recombination operators over this language. Ultimately, we plan to experiment with applying these operators intelligently, using heuristical and AI methods.

Our sequence mutation operators include:
\begin{itemize}
    \item \textbf{Phrase+:} Inserts a random phrase (token ngram) from the vocabulary set into the sequence at a random point. The phrase is sampled from a distribution slightly favoring more frequent phrases. This operator has the effect of adding a new phrase to a random clause.
    \item \textbf{Clause+:} Inserts a zero separator at a random point in the sequence. This has the effect of either splitting an existing clause into two, or creating a new empty clause.
    \item \textbf{Swap:} Transposes two elements in the sequence with each other. The separation between the elements is sampled from an exponential distribution, making long-distance swaps less likely. Changes to the ordering of phrases within a clause are silent mutations, but move phrases towards zeros, eventually swapping clauses to change meaning.
    \item \textbf{Negate:} Negates a random term within a clause.
    \item \textbf{Simplify:} Deletes duplicated terms within a clause.
\end{itemize}

Our recombination operators include:
\begin{itemize}
    \item \textbf{Crossover:} Combines two sequences by cutting each at a random point and concatenating the first half of one with the second of the other. The cut-points are selected from a distribution which favors cutting queries at or near zeros (clause boundaries). Cut points can also fall at the beginning or end of a query.
    \item \textbf{Swatch Insertion:} Inserts a piece of one query into a second. Using the cut-point selection defined above, we make two cuts in one sequence and one in a second, then insert the middle segment of the first between the two halves of the second.
\end{itemize}

\section{Preliminary Evaluation}
\subsection{Query Optimization}
In order to test the efficacy of this system, and to gather data for another project, the genetic algorithm presented above was used to develop a query for retrieving tweets relevant to predicting or understanding roadway traffic. Because of the subtle nature of the application and because our algorithm is still under development, human judgement is a necessary component of this task.

To use the query optimizer in semi-automated fashion, it is first given an initial population of human-designed queries to evolve. In practice, it is best if these are simple, one- or two-clause queries which the algorithm can refine, expand, and recombine. Because the algorithm is trained on limited data, and does not yet have a more general NLP or SAT-solver-based reasoning framework, the clauses of the queries it generates tend to accumulate low-quality terms, in addition to the useful ones. For this reason, a mild selective pressure against length is introduced, and every few hundred generations the evaluation is paused so that a few samples from the population can be manually simplified and improved. These samples are then fed back into the population for the algorithm to learn from.

After repeating this process for many generations, the below query was developed for identifying traffic-related tweets. Note that each OR clause tends towards a unique function within the query--selecting for tweets that match specific criteria. This is largely the action of the algorithm. The computer also contributed to the diversity of synonyms in clause 2, identified the utility of searching for numbers and directions to find highly specific tweets in clause 3, and recommended several stopwords common in irrelevant tweets in clause 4.

\begin{center}
    \includegraphics[width=2.5in]{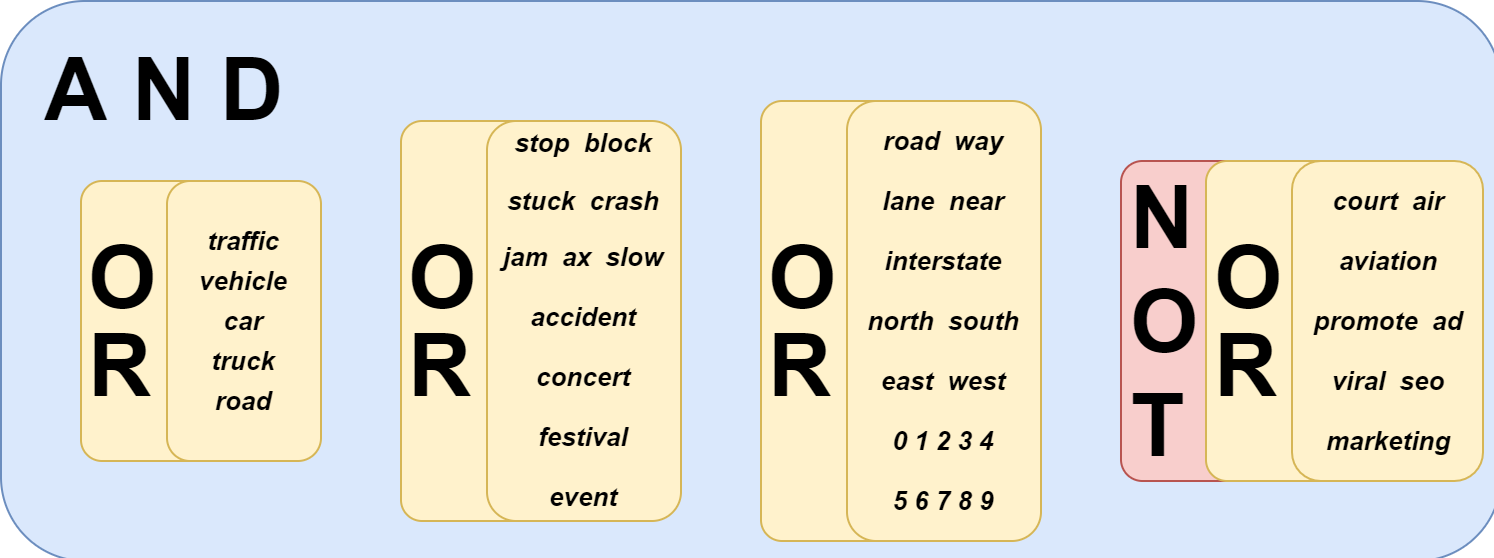}
\end{center}

\subsection{Sample evaluation}
With this query, we collected this dataset\footnote{\url{https://docs.google.com/spreadsheets/d/1PIm96ipZzqcE95f1eeBe814RddpErXaPP6e7lBFraEM/edit?usp=sharing}}, which includes 5000 traffic-related tweets from the region around Louisville in Spring 2020. As an evaluation, we manually classified 1000 of them (those from March 20 - April 20). First, we evaluated their relevance to traffic prediction. In order to be considered relevant, a tweet had to concern a recent, identifiable traffic pattern or incident:

\begin{center}
    \includegraphics[width=2.5in]{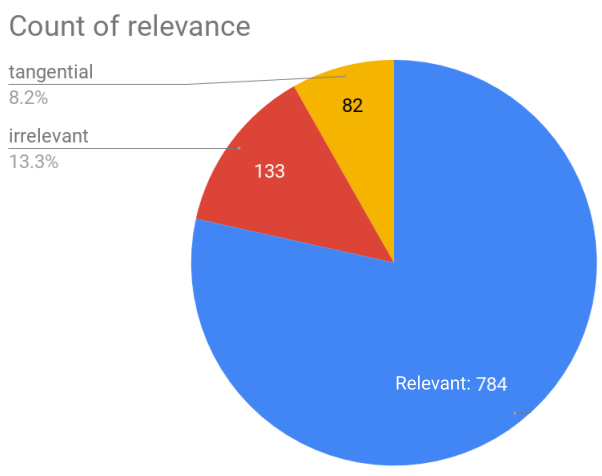}
\end{center}

We filtered out as tangential the tweets discussing traffic and accidents in the abstract, tweets describing an event that occurred far away or long ago, and tweets about events that were traffic-related, but would not impact other road users. Then, the relevant tweets were classified by the source they came from:


\begin{center}
    \includegraphics[width=2.5in]{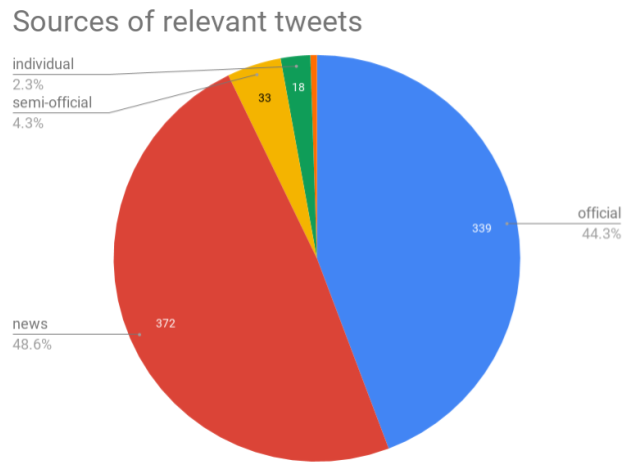}
\end{center}

The vast majority of traffic-relevant tweets come from official government and news sources. This makes sense, as these organizations have networks of human observers and remote sensors to detect these events, and a mandate to publish them quickly and accurately. By comparison, the traffic-related tweets posted by private individuals or organizations are far less frequent, and contain less precise information.



Moving on to examining the content of the tweets, we first categorized the relevant tweets based on the type of message they convey. A simple observation of an event during or after its occurrence is a report (i.e. '\#accident on bardstown'). An advisory makes a recommendation to the reader (i.e. '\#blackice! avoid newburg bridge'). The third type are plans for future events that may impact traffic (i.e. 'just about to leave for the big game! \#cats').

\begin{center}
    \includegraphics[width=2.5in]{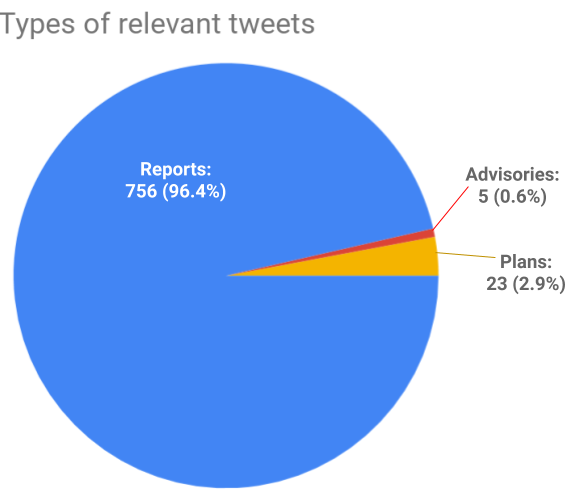}
\end{center}

Somewhat surprisingly, reports vastly outnumber advisories and plans. Advisories tend to be posted only by official sources, and then only when a change in public behavior is necessary for safety. Officials are probably cautious about the unintentional traffic problems their advisories might cause in other areas as drivers redirect. Plans seem to be rare in our dataset because they are almost exclusively posted by individuals and organizations, and because the impact on traffic is often not obvious or made clear in the text of the post.

Finally, we explored the set of relevant reports in more detail, classifying the reports based on the kind of traffic-related event they describe:

\begin{center}
    \includegraphics[width=2.5in]{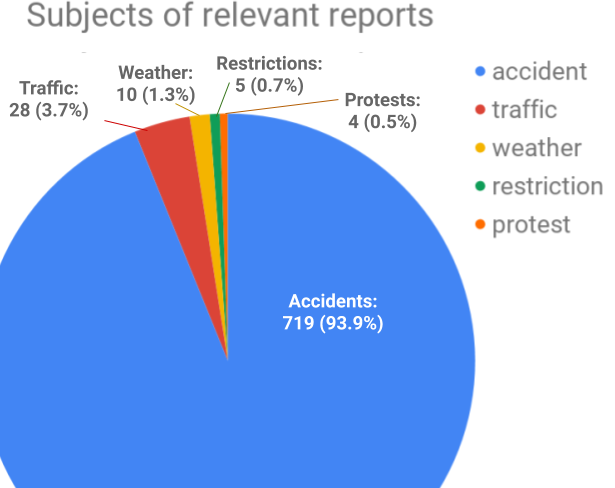}
\end{center}

The vast majority of reports concern accidents or crashes, often leading to traffic jams and lane restrictions. The next most common category were reports of traffic jams without a listed cause, followed by weather-related incidents, road closures and lane restrictions not related to accidents or weather, and finally there were four reports of protests interrupting traffic in our dataset.

Overall, 71.4\% of the tweets returned by our latest query contain relevant reports of traffic accidents and crashes. Previous attempts to collect data using either simple queries or more complex manually generated ones showed lower selectivity, as shown below.

\begin{figure}[H]
\centering
    \includegraphics[width=2.5in]{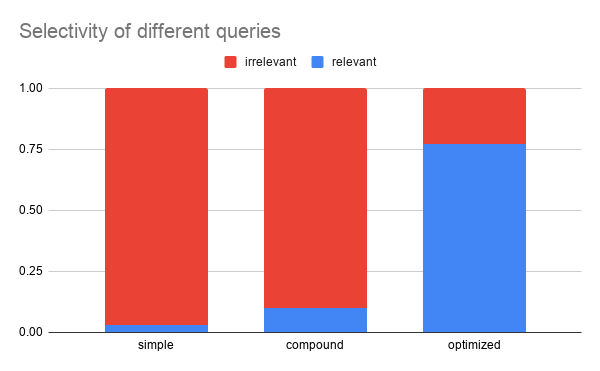}
   \vspace{-5mm}
  \caption{Percentage of tweets relevant to traffic sensing and prediction returned by various queries}
  \label{fig:tweet_relevance}
\end{figure}

\section{Data Extraction}

As a trial, we performed manual data extraction on a sample of 100 accident report tweets (a subset of the 714 identified earlier). My goal was to determine for each tweet where and when the reported accident occurred.

Most of the reports did not contain any specific information about the time the accident in question occurred. Instead, we tried to deduce from context and word choice the approximate recency of the accident vs the time the tweet was posted. About 91\% of posts seemed to be about very recent accidents, probably less than one hour before they were posted. Needless to say, this is a very qualitative process and getting a computer to perform the same deduction will be challenging.

Less than 1\% of tweets are geotagged, but about 85\% of reports include specific information about where an accident took place embedded in the text content or attached images. These geo-specific reports are mostly from official and news sources; the remainder simply indicate that an accident occurred within a city or area, but without enough information to specifically identify the incident. Reports that do specify the location tend to do so by referencing a business or intersection for street accidents (i.e. ``pileup in front of chili's'' or ``ax on bardstown rd at rose st''), or a ramp or mile-marker for interstate and highway accidents (i.e. ``crash on I-264 at exit 103 for poplar level.'' or ``I-64: inverted semi at mm23.7''). Eventually, we will want to use natural language processing techniques to automatically identify the relevant data within the tweet; for now it is donemanually.


Finally, we loaded this GIS data into Google Earth, to produce the map below:

\begin{center}
    \includegraphics[width=2.5in]{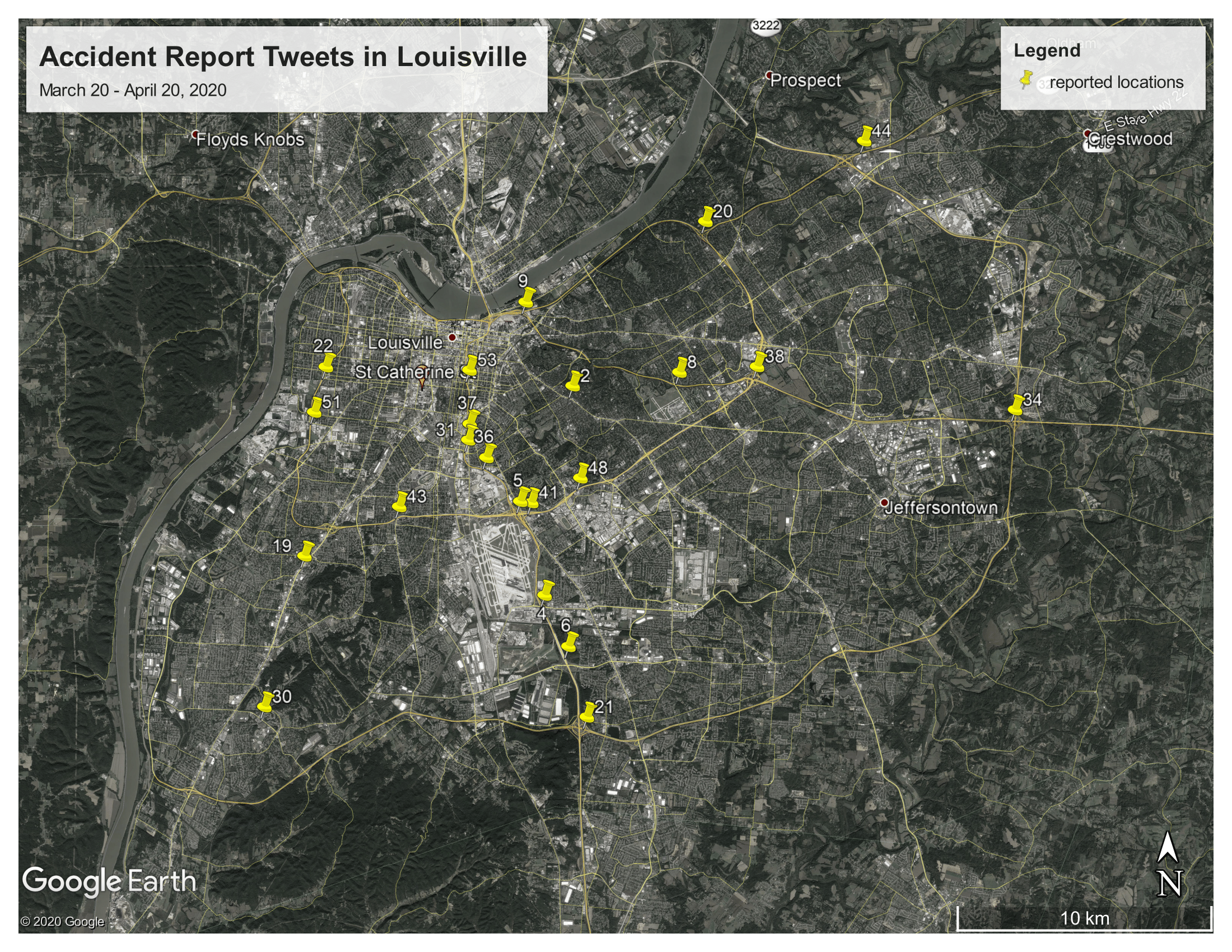}
\end{center}

\bibliography{refs}

\begin{thebibliography}{1}

\bibitem{tamine2003multiple}
L.~Tamine, C.~Chrisment, and M.~Boughanem, ``Multiple query evaluation based on
  an enhanced genetic algorithm,'' {\em Information Processing \& Management},
  vol.~39, no.~2, pp.~215--231, 2003.

\bibitem{losee1988integrating}
R.~M. Losee and A.~Bookstein, ``Integrating boolean queries in conjunctive
  normal form with probabilistic retrieval models,'' {\em Information
  processing \& management}, vol.~24, no.~3, pp.~315--321, 1988.

\bibitem{safari2019optimizing}
K.~Safari and S.~Sanner, ``Optimizing search api queries for twitter topic
  classifiers using a maximum set coverage approach,'' {\em arXiv preprint
  arXiv:1904.10403}, 2019.

\end{thebibliography}
\bibliographystyle{ieeetr}

\end{document}